\documentclass[creativecommons]{eptcs}
\providecommand{\event}{F-IDE-18} 
\usepackage{breakurl}
\usepackage{underscore}           
\usepackage{graphicx}
\usepackage{subcaption}
\usepackage{enumitem}
\setlist[itemize]{leftmargin=*}

\title{Integrating User Design and Formal Models within PVSio-Web}
\author{Nathaniel Watson
\institute{Department of Computer Science\\
University of Waikato, Hamilton, New Zealand}
\email{nathaniel.nw@gmail.com}
\and
Steve Reeves
\institute{Department of Computer Science\\
University of Waikato, Hamilton, New Zealand}
\email{steve.reeves@waikato.ac.nz}
\and
Paolo Masci
\institute{High Assurance Software Laboratory (HASLab)\\
INESC TEC and Universidade do Minho, Braga, Portugal}
\email{paolo.masci@inesctec.pt}
}

\def\titlerunning{Integrating User Design and Formal Models within PVSio-Web}
\def\authorrunning{N. Watson, S. Reeves \& P. Masci}

\begin{document}

\maketitle

\begin{abstract}
Creating formal models of interactive systems has wide reaching benefits, not only for verifying low-level correctness, but also as a tool for ensuring user interfaces behave logically and consistently. Despite this, tools for designing user experiences and tools for creating and working with formal models are typically distinctly separate systems. This work aims to bridge this divide by allowing the generation of state machine diagrams and formal models via a simple, interactive prototyping tool that mirrors the basic functionality of many modern digital prototyping applications.
\end{abstract}

%


\section{Introduction}\label{sec:intro}
Formal (i.e., mathematical) methods have a wide range of uses across the software development process, from providing a mechanism for specifying and analysing early-stage requirements through to being used as a basis for testing and reasoning about implemented systems. 

Despite their benefits, formal techniques have not seen widespread adoption outside of software projects and processes that involve safety-critical systems. A number of reasons have been suggested as a cause for this~\cite{kreiker2011manifesto}. An important barrier that is usually overlooked relates to the usability of the tools. Developers of formal methods tools are usually the main end users of the tool, resulting in tool front-ends that are tailored to one person's taste and skills (or a small team at best). Other barriers include lack of functionalities for integrating formal methods tools within other less formal tools and workflows adopted by developers. We refer the reader to~\cite{bowen-models-2007,jaspan2009software,larsen2010formal,paiva2007reverse,whittle2014state} as examples of the discussion in this area.

With the current popularity of web technologies for creating prototypes (see Section~\ref{sec:subtraction}), it is perhaps an area that could be re-explored in the context of modelling early stage informal design artefacts, either as part of a model-driven engineering approach or to support the use of more formal methods within a development process. 

This paper presents our work on extending an existing formal methods tool, PVSio-web~\cite{cav15}, with a new front-end that can be used by designers who have no experience or training in formal concepts. PVSio-web is an open-source toolkit for model-based development of critical interactive systems. It aims to support a multi-disciplinary team of developers who are using PVS~\cite{pvs} as main modelling and verification technology. The toolkit integrates specialised front-ends specifically designed for different target users. The original PVSio-web toolkit includes three main front-ends:

\begin{itemize}[topsep=0pt,noitemsep]
\item A {\em Prototype Builder} allows user interface experts to create high fidelity prototypes of interactive systems. The visual aspect of the prototype is based on a picture of the device. The behaviour of the prototype
is an executable formal model created by formal methods experts.

\item A {\em Simulator} executes the interactive prototype. The visual appearance of the prototype is rendered in a Web browser. The logic of operation of the prototype is executed in PVSio~\cite{munoz2003rapid}, the native component of the PVS system for animating PVS models.
User actions over input widgets (e.g., button presses) are translated by the Simulator into PVS expressions that can be evaluated in PVSio.
The result of the PVSio evaluation is rendered in the Web browser using the output widgets of the prototype, 
so that the visual appearance of the prototype closely resembles that of the real system in the corresponding states.
Formal methods experts can use this front-end to discuss analysis results to domain experts.

\item An {\em Emucharts Editor} provides software engineers with a visual modelling language for specifying the behaviour of the system as a visual diagram based on the stateflow notation. The editor provides functionalities for automatic translation of the visual diagram into an executable formal model.
\end{itemize}

\smallskip\noindent
{\bf Contribution.~}
A new PVSio-web front-end, the {\em PIM Prototyper}, is developed for creating interactive mock-up prototypes based on a series of sketches representing the different screens of the prototype. The front-end mimics the typical characteristics and features of popular digital design tools commonly used by developers who have no prior experience or training in formal concepts. The behaviour of the prototypes developed with the PIM Prototyper can be automatically translated into a formal model that can be further analysed by formal methods experts. The new front-end can be downloaded as part of the latest distribution of PVSio-web\footnote{\url{http://www.pvsioweb.org}}.

\smallskip\noindent
{\bf Organisation.~} The rest of the paper is organised as follows. Section~\ref{sec:subtraction} provides an overview of the typical features offered by popular design tools for creating mock-up prototypes. Section~\ref{sec:pims} identifies a formal modelling language suitable for representing the behaviour of mock-up prototypes. Section~\ref{sec:reqs} identifies design requirements for the new front-end. Section~\ref{sec:dev} presents our implementation of the front-end. Related work is discussed in Section~\ref{sec:related}. Finally, Section~\ref{sec:conclu} concludes the paper.

\section{Popular design tools for creating mock-up prototypes}\label{sec:subtraction}
Little data exists on the popularity of prototyping tools for creating mock-up prototypes based on a series of sketches.
The 2015 Subtraction.com Design Tools Survey\footnote{\url{http://tools.subtraction.com}} provides a good starting point for determining which prototyping tools are commonly used by developers. It shows that web technologies (HTML and CSS) are by far the most popular toolkit for creating prototypes. The survey identifies also four tools commonly used by developers, which are discussed in the rest of this section.

{\em InVision}~\footnote{\url{https://www.invisionapp.com}} is one of the most popular prototyping tools. The tool allows developers to create annotated mock-ups, and provides a preview mode for navigating the prototype and perform basic user testing around the prototype. There are no features present for automatically analysing any aspects of a prototype, with the focus instead seeming to be on user-provided feedback, such as comments and annotations. The popularity of this tool is also shown by \cite{oreilly2016survey}, which lists it as one of the three dominant tools used by the survey's respondents.

{\em Marvel}~\footnote{\url{https://marvelapp.com}} is a web application that provides similar functionality to InVision. No high-level representation or visualisation of the links between screens exists. Likewise, there is no sort of analysis of the design or navigation structure. Much like InVision, the focus seems to be on allowing potential users of the final system to provide written feedback via comments and annotations. 

{\em Fluid UI}~\footnote{\url{https://www.fluidui.com}} differs from the previous two tools in that it provides tools for creating detailed mock-ups as the starting point for a prototype, rather than using static images. While this does allow a range of different widget types to be added to the prototype, this does not have any bearing on how the widgets can be interacted with --- the differences are purely visual. However, because the widgets are more than just images, links between screens are associated with the widgets themselves rather than hotspots drawn over them. 

{\em Axure}~\footnote{\url{https://www.axure.com}} is a desktop application for creating mock-ups and prototypes. Like Fluid UI, the integration of mock-ups and prototyping into a single environment means hotspots are not used, but instead interactions are associated with components of the mock-up. The mock-up components are backed by HTML and hence their types (e.g., input, output or any other more specific type) are tightly linked to the component and it is often not necessary for the user to explicitly specify the general type of the component. It has features that make it possible to create an approximation of the actual system behaviour using the graphical menus and logic representations provided by the tool.

The four tools described above fall into two clear categories: those focused solely on facilitating the creation of user interface prototypes, and those that provide prototyping as an extension of mock-up creation functionality. With their focus on the use of hotspots and static images, InVision and Marvel fall into the former category, while the latter contains Axure and Fluid UI, with their prototyping tools being based on the detailed components that make up the screens of the underlying mock-up. 
None of the tools provides any sort of automated analysis of the prototype, highlighting an area where designers, in addition to formal methods practitioners, could benefit from the creation of models that are able to be used to generate this sort of information. All the tools provide a preview mode, through which the actions defined by the designer could be triggered by a user directly interacting with the prototype.

\section{Formal modelling of mock-up prototypes}\label{sec:pims}
Presentation Interaction Models (PIMs) are a type of formal model intended for representing interactive systems. They are made up of two main components: the PIM and the PM (Presentation Model). A PM contains information about a single user interface state, primarily the widgets (input and output elements) it contains. Each widget can be associated with behaviours that they either trigger or respond to, depending on the widget type. They can take two forms: \textit{I-behaviours}, which represent changes in the user interface's state, and \textit{S-behaviours}, representing functionality of some sort in the underlying system. A PIM is a finite state automaton for which each state is a PM and each transition is an I-behaviour that can be triggered from the source PM.

This structure is particularly well-suited to the formalisation of informal design artefacts, e.g.,~\cite{bowen-models-2007} demonstrates how a paper prototype --- sketches of a user interface's screens made on paper --- can be used to direct the creation of PIMs and PMs. This allows existing methods, both informal (for example using paper-prototypes as part of a wider user-centric design process) and formal to be used in unison. This established ability to integrate design artefacts with PIMs makes them particularly well-suited to this work. Modern digital prototyping tools can be considered an evolution of paper prototypes and so many of the potential uses of this sort of pairing that are described by \cite{bowen-models-2007} are also applicable to this work. The tool PIMed\footnote{\url{https://sourceforge.net/projects/pims1}} supports a design using PIMs.

\section{Design Requirements}\label{sec:reqs}
Design requirements of the tool being created in this work have been identified based on the research described in Sections~\ref{sec:subtraction} and~\ref{sec:pims}. The aim is to create a tool that allows users (as developers) to build on their knowledge and previous experience with informal prototyping tools. At the same time, the tool should also provide functionalities to automatically generate a formal model representing the behaviour of the mock-up prototype --- the formal model will feed subsequent analyses carried out by formal methods experts. Four essential features have been identified:
\begin{itemize}
\item \textbf{Multiple screens}. The tool must support adding a unique image for each screen and navigating between the screens within the editing interface. These correspond to states within the generated PIMs.
\item \textbf{Ability to add hotspots to screens}. It must be possible for each screen to have a set of hotspots associated with it. Furthermore, developers should be able to create hotspots by drawing them over the screen's image. Each hotspot will correspond to a single widget within the PIM and belong to their parent screen's state.
\item \textbf{Ability to link hotspots}. The prototype only becomes useful if some sort of interaction is possible. Allowing hotspots to be used as a way to navigate between screens was the primary approach used by all tools in Section~\ref{sec:subtraction}, and so similar functionality should be possible via the new tool's interface. A link between two screens will correspond with a transition within this PIM. In particular, they will be represented by transitions in the {\em I-behaviours}, which describe changes in the state of the user interface. It is also planned that {\em S-behaviours} should be able to be specified, although this would primarily be for the sake of creating more detailed models: any use within the prototyping tool's viewing mode will be nothing more than simple visual effects (for example, highlighting affected hotspots when the user triggers an {\em S-behaviour}) rather than attempting to simulate the actual functionality that may be associated with the behaviours.
\item \textbf{Viewing mode}. This facilitates basic manual evaluation/debugging and user testing of the prototypes the tool will produce. Unlike the other core features, this does not correspond to any aspect of PIMs. Rather, the inclusion acts to demonstrate that the tool is not simply an alternative way of generating an interface model, but a multi-purpose environment that supports multiple evaluation approaches (model-based analysis and user-based testing).
\end{itemize}

It is envisioned that the tool being created should be able to be used both as an alternative for the core prototyping functionality of tools such as InVision and Marvel, and as an intuitive way of creating PIMs. However, expecting it to be able to completely replace the larger commercial prototyping tools within a design and development workflow is unrealistic: the tools described in Section~\ref{sec:subtraction} generally offer a number of features that act to augment basic prototype creation, such as commenting and simple animation. Because these features do not affect the structure of the prototype itself, the tool being created can still be considered an accurate representation of the plausibility of combining informal prototyping and formal modelling.

\section{Development of the new prototyping front-end}\label{sec:dev}
Design requirements were specified through the use of a visual mock-up representing the screens of the front-end, and a series of \textit{user stories} intended to provide a view of a system's features from the view of a user. The particular type of user stories adopted are based on those described by~\cite{north-story-2007} and provide a short description of the user's goal, the steps that should be followed to complete a task and the expected outcome of following the steps. This structure was chosen as, in addition to providing an unambiguous reference when implementing features, it is well-suited to being adapted into test cases for automated testing of the developed front-end. Once implemented, these tests can both act as a check for what has been completed and ensure that the feature set remains consistent in any future development.

\begin{figure}[t]
\centering
    \begin{subfigure}[b]{0.48\textwidth}
        \includegraphics[width=\textwidth]{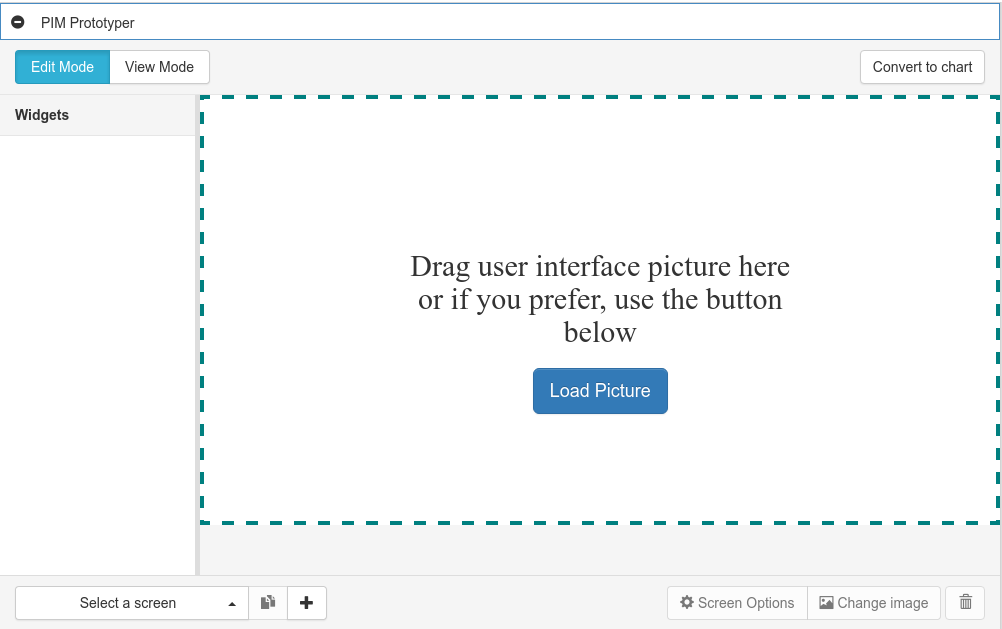}
        \caption{The main PIM prototyping interface.}
        \label{fig:pimp-initial}
    \end{subfigure}
    \begin{subfigure}[b]{0.48\textwidth}
        \includegraphics[width=\textwidth]{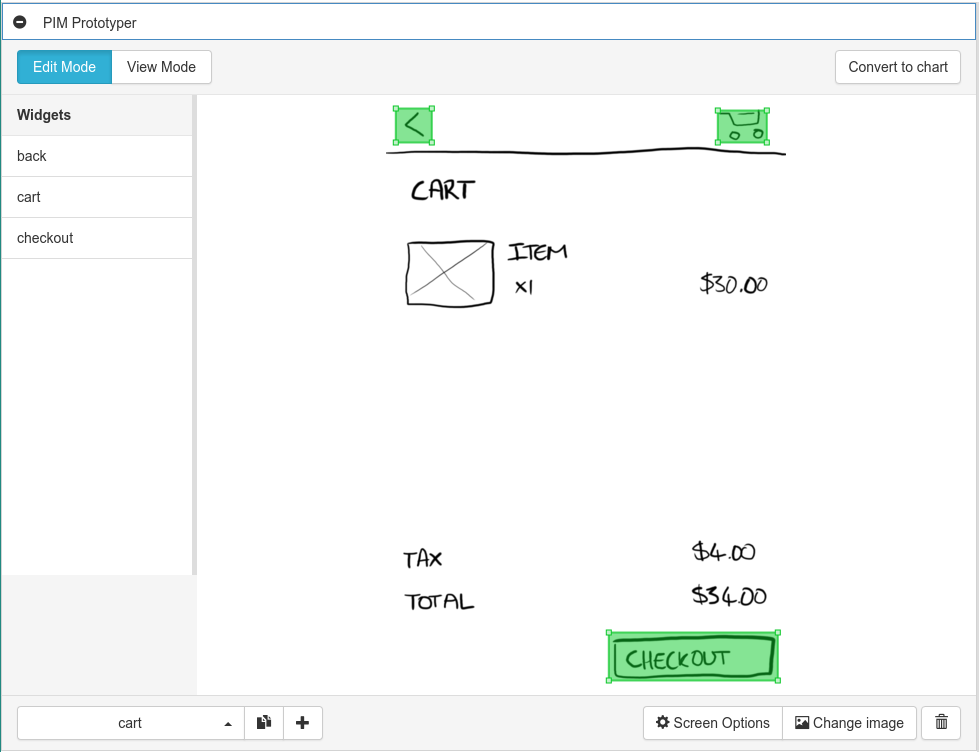}
        \caption{A prototype screen being edited.}
        \label{fig:pimp-screen}
    \end{subfigure}
    
    \begin{subfigure}[b]{0.48\textwidth}
        \includegraphics[width=\textwidth]{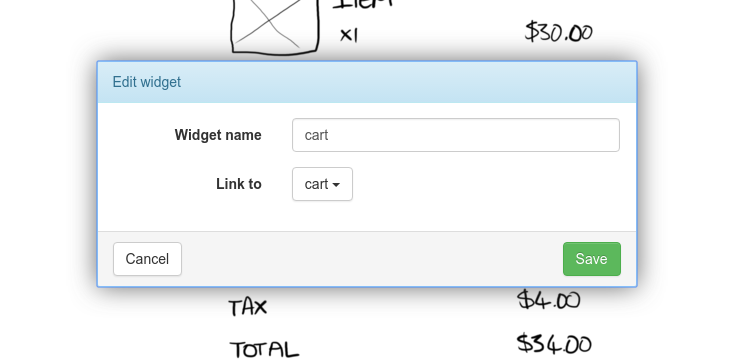}
        \caption{The options for configuring a widget.}
        \label{fig:pimp-edit}
    \end{subfigure}
    \begin{subfigure}[b]{0.48\textwidth}
        \vspace*{8pt}
        \includegraphics[width=\textwidth]{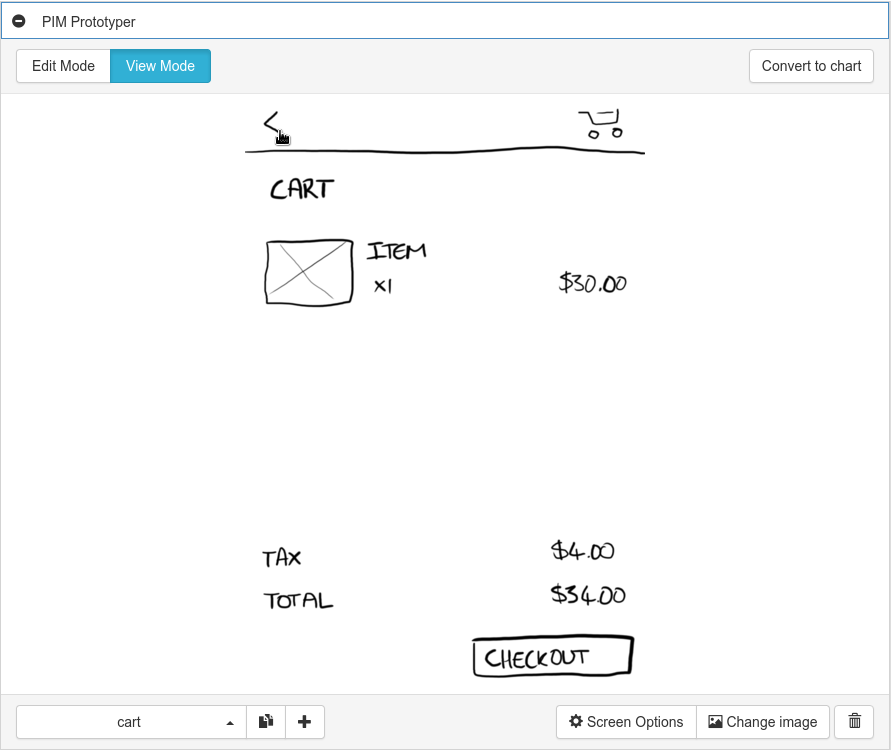}
        \caption{A prototype in the interactive viewer mode.}
	    \label{fig:pimp-view}
    \end{subfigure}
    \caption{Screens of the PIM Prototyper.}\label{fig:pimeditor}
\end{figure}

\begin{figure}[t]
	\centering
		\includegraphics[width=0.8\textwidth]{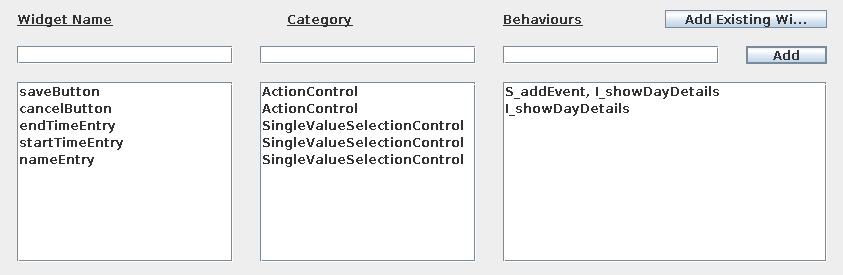}
	\caption{The widget configuration screen in PIMed.}
	\label{fig:pimed-widgets}
\end{figure}

The developed extension provides a PIM prototyping mode that can be accessed as a module within PVSio-Web (see Figure~\ref{fig:pimp-initial}). This makes it completely optional when using the other tools present within the PVSio-Web environment, and likewise, it can be used without needing to activate any of the other tools. This contributes towards the aim of making a tool that can be used by people without any knowledge of formal modelling, as there is no need to interact with or configure anything outside the PIM Prototyping module. 
Once activated, the module presents the main prototype editing interface, shown in Figure \ref{fig:pimp-initial}. 
In the case of the PIM prototyping tool, loading an image will create a new screen within the prototype. Subsequent screens can be added from the bottom menu bar, which also provides functionality for switching between screens and modifying their settings, such as selecting a different image to use and setting the initial screen.
Adding hotspots to screens follows the same paradigm that was seen to be common in the commercial tools looked at in Section~\ref{sec:subtraction}: dragging over the image will create a new hotspot and display it as a rectangle over the image (see Figure~\ref{fig:pimp-screen}). The hotspot can then be configured (see Figure~\ref{fig:pimp-edit}). The options here were intentionally kept simple. As was seen in Section \ref{sec:subtraction}, most prototyping tools offer little more than the ability to set which screen a widget links to. This contrasts with the widget configuration screen within PIMed, shown in Figure~\ref{fig:pimed-widgets}. Excluding the widget's name, each widget only has two pieces of information: a category and the behaviours associated with it. However, the way this is presented in PIMed is quite technical, requiring knowledge of the expected syntax for behaviour names, the meaning of \textit{S} and \textit{I} behaviours and understanding the various categories that a widget might belong to. For a tool aimed at users with an existing understanding of PIMs these are all reasonable expectations, but with the tool that has been created aiming to lower the barrier of entry for people inexperienced with formal modelling, these details may be overly complex, particularly if their primary goal is to create an interactive prototype, not a formal model. As such, the widget editing screen aims to provide a way for some of this information to be specified in a way that reflects what the user intends for a widget to do, rather than directly reflecting how it would be represented in the modelling syntax. 

As discussed in Section~\ref{sec:reqs}, the ability to interact with a simple simulation of the prototype is a core feature and use-case of existing prototyping tools. The implemented viewer mode displays the image for the active screen and allows the hotspots defined within the screen to be clicked by the user. Upon clicking a hotspot, the displayed image and hotspot regions switch to those belonging to the screen the clicked hotspot links to. In this way it is possible to move between screens of the prototype and gain an understanding of how it functions. This could also be used to perform user testing on the prototype, in the same way existing prototyping tools are. 

Unlike many existing prototyping tools, the viewer mode is not distinctly separate from the edit mode (for example, accessing the preview mode in Marvel opens in a new window and requires waiting for the interactive prototype to load). Instead, the editing and viewing modes can be quickly switched between, making it possible to interactively test the prototype until an issue is found, switch to editing the screen, then switch back to interacting after making any necessary changes.

\subsection{Integration with existing PVSio-Web features}\label{sec:integration}

\begin{figure}[t]
	\centering
	\includegraphics[width=0.75\textwidth]{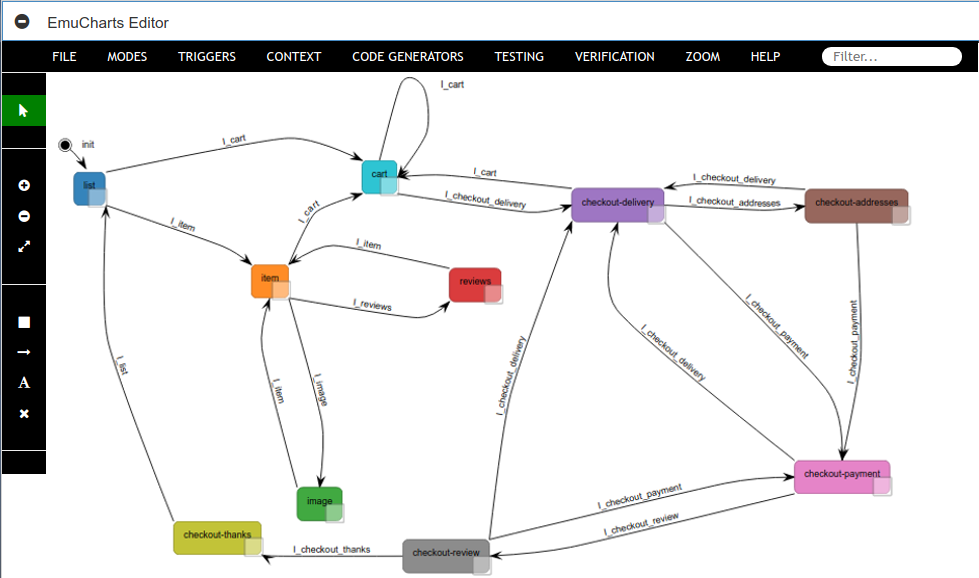}
	\caption{PIM generated from the prototype, visualised as a state machine in the Emucharts Editor.}
	\label{fig:emuchart}
\end{figure}

Once a prototype has been built, the \textit{Convert to chart} button shown in the top-right of Figure~\ref{fig:pimp-screen} can be used to automatically generate a PIM model. The model can be visualised as a state machine diagram within PVSio-Web's Emucharts Editor, which allows formal methods experts to perform further editing of the model (see Figure~\ref{fig:emuchart}). In this way all PIM-related functionalities are incorporated in a formal model that can be imported in the PVSio-web Prototype Builder and later used to create high fidelity prototypes. This includes also the ability to generate abstract test cases for the user interface described by the PIM (described by~\cite{nathan-honours}). When used in conjunction with the prototyping tool, this could allow a standard prototype-based design workflow to be followed by a user interface designer, while still giving a developer implementing the interface an unambiguous, concise reference for its core functionality (whether just as a reference or as a base for more detailed, executable test cases).

In order visualise a PIM as a state machine diagram, some assumptions are made about the information provided via the PIM Prototyper. The main assumption is regarding the category of the widgets being represented. Each widget within a PM can be associated with a category that provides information regarding how it behaves and the functionality it provides~\cite{BowenThesis2008}. Since the prototyping tool is based on widgets that accept user input (i.e., clicking), it is reasonable to assume that they all fit within the \textit{Event Generator} category. Furthermore, given that each widget triggers a change in the overall interface state, it has also been assumed that the widgets are not \textit{Value Selection} or \textit{Binary Selection Controls}, but rather fit within the \textit{Action Controls} specialisation of the category hierarchy. Beyond this, the conversion from prototype to state machine is straight-forward: each screen within the prototype becomes a state, with the widgets within each screen mapping directly to the widget information stored by each state. Transitions between states are created from the inter-screen links associated with widgets within the prototype. In cases where the same screen is linked to by multiple widgets within a screen, the resulting transitions are automatically merged into a single transition.

\section{Related work}\label{sec:related}
Prototyping and model animation has been used by others as a means to facilitate validation of formal models and formal analysis results with domain experts and engineers. An example is the SCR toolset developed by Heitmeyer et al.~\cite{heitmeyer1998scr}, which includes functionalities for creating interactive prototypes based on formal specifications. Another example is Ladenberger's et al.~\cite{hoang2016validating,ladenberger2009visualising} B-Motion Studio for animating Event-B models, and Oda's et al.~\cite{vdmpad,oda2015vdm} toolkit for animating executable VDM models using interactive mock-up prototypes and ad-hoc visual diagrams. These and other similar tools are meant to be used by formal methods experts to engage with team members that are not familiar with formal methods technologies --- a formal model is developed first by formal methods experts, and the tool is used to demonstrate the behaviour of the model to non-experts of formal methods. The tool presented in this paper is designed for a different workflow --- non-experts of formal methods use our tool to create a mock-up prototype based on a series of inter-linked sketches of user interface screens, and to convert the prototype into a formal model that can be further analysed by formal methods experts.

Other approaches to integrating formal methods technologies in product development focus on the use of visual diagrams and domain-specific notations. Winckler and Palanque~\cite{winckler2003statewebcharts}, for example, introduced statecharts modelling extensions that could ease modelling and analysis of web apps. Acerbis et al.~\cite{Acerbis2015} extended the modelling language of an existing model-driven development environment to facilitate the adoption of verification technologies in the context of mobile apps. A similar approach was also used by Grov et al.~\cite{grov17,Grov2018} to lower the knowledge barriers for using sophisticated features of verification tools. In our work, the problem is addressed from a different angle: sketches of user interface screens and inter-links between screens are used {\em as} visual diagrams to specify the behaviour of the system.

\section{Discussion and Conclusion}\label{sec:conclu}
This work set out with a goal of exploring the possibility of merging formal model creation with popular approaches to informal interface design. Existing research suggested that this sort of integration could help in increasing the use of formal methods by making them more accessible to users inexperienced with formal concepts and more easily integrated with existing workflows. The solution that has been created demonstrates that such a merging of existing formal and informal tooling and techniques is certainly possible. The user-facing features of the resulting tool are based heavily on the core features of the most widely used interface prototyping tools, providing a result that allows user interface prototypes to be created and interacted with in a manner that is familiar to many designers. At the same time, the tool's underlying basis in PIMs and PMs, made possible through the integration with features in the PVSio-Web environment, has been shown to allow the generation of accurate models, potentially allowing model-based techniques (such as automatic test generation) to be used and benefited from within the wider development process, without extra effort being needed to manually create the formal models.

The outcome is not without its shortcomings: this work was carried out as part of the thesis of the first author, and some of the planned features could not be implemented due to time constraints, and the prototyping tool is not able to represent or generate the full set of features offered by PMs. While this does perhaps limit the broad usefulness of the current version of the tool as a means of generating and simulating detailed models, it is hoped that future work will be able to explore ways in which these missing features could be integrated with the tool. The development approach that was followed --- structuring the code to follow standard architectural patterns and providing extensive unit and behavioural tests --- should allow future work to build on what has been created with relative ease and confidence that the existing features continue to function correctly.

The main areas that future development may want to consider would be implementing the features that were planned but not completed. This would involve adding tighter, two-way integration between the PIM Prototyping tool, the Prototype Builder, and the Emucharts Editor, as well as providing basic means of defining and representing {\em S-behaviours} within a mock-up prototype. Beyond this, more exploration could be performed into means for describing expected properties of the prototype, e.g., {\em ``screen X can be reached only by authenticated users''}. These properties could be analysed either by interacting with the prototype, or using formal methods technologies. In order to maintain the goal of making formal modelling tools more familiar (and therefore more accessible), any attempt at adding these types of features would need to involve considering and reflecting how popular existing tools (outside of formal modelling) make use of similar information.

On a broader scale, the prototyping tool could potentially be extended to directly provide model-based information and analysis to a user as they are creating a prototype. This might involve automatically analysing the model corresponding to the prototype in order to identify issues such as states becoming unreachable, allowing a user to rapidly make improvements to their prototype based on the immediate feedback. A different direction would be to allow S-behaviours within a prototype to be associated with actions in a PVS specification, allowing simulation of more detailed functionality within a prototype. This would not be of any benefit to the goal of making formal model creation more accessible, but would place the PIM prototyping mode on-par with PVSio-Web's original prototyping mode in terms of its ability to simulate existing models.

It may also be beneficial to run a usability study of some sort on the tool that has been created. While it has been assumed that basing the tool's features on the core features provided by existing tools should provide a familiar, easily accessible experience, our work does not definitively confirm the validity of any such assumptions. A user study involving the tool being used in realistic situations with users who are not familiar with formal modelling concepts would allow the validation of these assumptions and may also reveal areas of the tool that could be improved further.

{
\smallskip\noindent
{\bf Acknowledgement.}
Paolo Masci's work is supported by ERDF - European Regional Development Fund through the Operational Programme for Competitiveness and Internationalisation - COMPETE 2020 Programme within project POCI-01-0145-FEDER-006961, and by National Funds through the Portuguese funding agency, FCT (Funda\c{c}\~{a}o para a Ci\^{e}ncia e a Tecnologia) as part of project UID/EEA/50014/2013.
}

\bibliographystyle{eptcs}
\bibliography{main}

\end{document}